\documentclass[manuscript]{acmart}

\renewcommand\footnotetextcopyrightpermission[1]{}

\usepackage{mathtools}
\usepackage{amsmath}
\usepackage{subfig}   
\usepackage{enumitem} 
\usepackage{booktabs}

\newcommand{\code}[1]{\texttt{#1}}
\DeclarePairedDelimiter{\ceil}{\lceil}{\rceil}
\DeclarePairedDelimiter{\floor}{\lfloor}{\rfloor}
\newcommand{\bigQ}[1]{\mathbf{Q_{#1}}}
\newcommand{\comm}[1]{\mathbf{t^{#1}}}

\begin{document}

%
\title{A Generalization of the Allreduce Operation}

%
\author{Dmitry Kolmakov}
\email{kolmakov.dmitriy@huawei.com}
\affiliation{%
  \institution{Central Research Institute, Huawei Technologies}
  \city{Moscow}
  \country{Russia}
}

\author{Xuecang Zhang}
\email{zhangxuecang@huawei.com}
\affiliation{%
	\institution{Central Research Institute, Huawei Technologies}
	\city{Hangzhou}
	\country{China}
}

%

%
\begin{abstract}
\textit{Allreduce} is one of the most frequently used MPI collective operations, and thus its performance attracts much attention in the past decades. Many algorithms were developed with different properties and purposes. 
We present a novel approach to communication description based on the permutations inspired by the mathematics of a Rubik's cube\texttrademark\  where the moves form a mathematical structure called group. Similarly, cyclic communication patterns between a set of $P$ processes may be described by a permutation group. This new approach allows constructing a generalization of the widely used \textit{Allreduce} algorithms such as Ring, Recursive Doubling and Recursive Halving. 
Using the developed approach we build an algorithm that successfully solves the well-known problem of the non-power-of-two number of processes which breaks down the performance of many existing algorithms. The proposed algorithm provides a general solution for any number of processes with the dynamically changing amount of communication steps between $\ceil{\log{P}}$ for the latency-optimal version and $2\ceil{\log{P}}$ for the bandwidth-optimal case.  
\end{abstract}



\maketitle

\section{Introduction}

Message Passing Interface MPI \cite{mpi-forum} is a de facto standard framework for distributed computing in many HPC applications. MPI collective operations involve a group of processes communicating by message passing in an isolated context, known as a communicator. Each process is identified by its rank, an integer number ranging from $0$ to $P-1$, where $P$ is the size of the communicator. 

\textit{Allreduce} is a complex collective operation that performs a combination of vectors owned by processes into a result vector which is distributed back to the processes. In MPI \textit{Allreduce} operation can be accessed by calling \code{MPI\_Allreduce()} function. Five-year profiling of HPC applications running in production mode \cite{Rabenseifner99} at the University of Stuttgart revealed that more than 17\% of the time was spent in the \code{MPI\_Allreduce()} function and that 28\% of all execution time was spent on program runs that involved a non-power-of-two number of processes. The average size of the data array involved in \textit{Allreduce} is about 425 Bytes which makes special demands on the \textit{Allreduce} performance shown for small data sets.

Recently one more application for the \textit{Allreduce} operation has emerged. Distributed training of Deep Neural Networks (DNN) \cite{DDL01, DDL02, DDL03, DDL04} uses \textit{Allreduce} operation to synchronize neural network parameters between separate training processes after each step of gradient descent optimization. This new application involves data sets of medium and big sizes which depends on a particular neural network model bringing new requirements to the \textit{Allreduce} performance.

The \textit{Allreduce} operation has been studied well in recent years, and many algorithms were proposed which behave differently depending on data size, network parameters and the number of processes. Ring algorithm \cite{Barnett93,Patarasuk09} performs \textit{Allreduce} optimally in terms of bandwidth, it is a so-called bandwidth-optimal algorithm, but requires a linear number of steps $2(P-1)$, so it is suitable only for big-sized tasks. Recursive Halving algorithm proposed in \cite{Rabenseifner04} is also bandwidth-optimal but can be done in a logarithmic number of steps $2\log(P)$ (here and further in the paper $\log(P)$ stands for $\log_{2}(P)$). Recursive Doubling algorithm \cite{Thakur05} performs the \textit{Allreduce} operation in the minimal possible number of steps - $\log(P)$, it is a so-called latency-optimal algorithm, but it has a considerable overhead in terms of bandwidth so it can be used effectively only for small messages. A hybrid approach allows combining Recursive Doubling and Recursive Halving to vary the number of steps from $\ceil{\log(P)}$ for the latency-optimal version to $2\ceil{\log(P)}$ for the bandwidth-optimal version. A general approach is to start the reduction phase with the bandwidth-optimal algorithm and when the data size to be exchanged becomes small enough to switch to the latency-optimal version. Thus the number of steps can be decreased by the cost of additional bandwidth overhead. Recursive Doubling, Recursive Halving, and Hybrid algorithms work only for the power-of-two number of processes and require additional work in the general case. 

We propose a generalization for the \textit{Allreduce} operation based on \textit{group theory} which enables the development of a generalized algorithm successfully solving the problem of non-power-of-two number of processes. Our solution works optimally for any number of processes and can vary the number of steps, trading off latency for bandwidth, even for a prime number of processes. 

\begin{figure}[h!]
	\centering
	\includegraphics[scale=0.15]{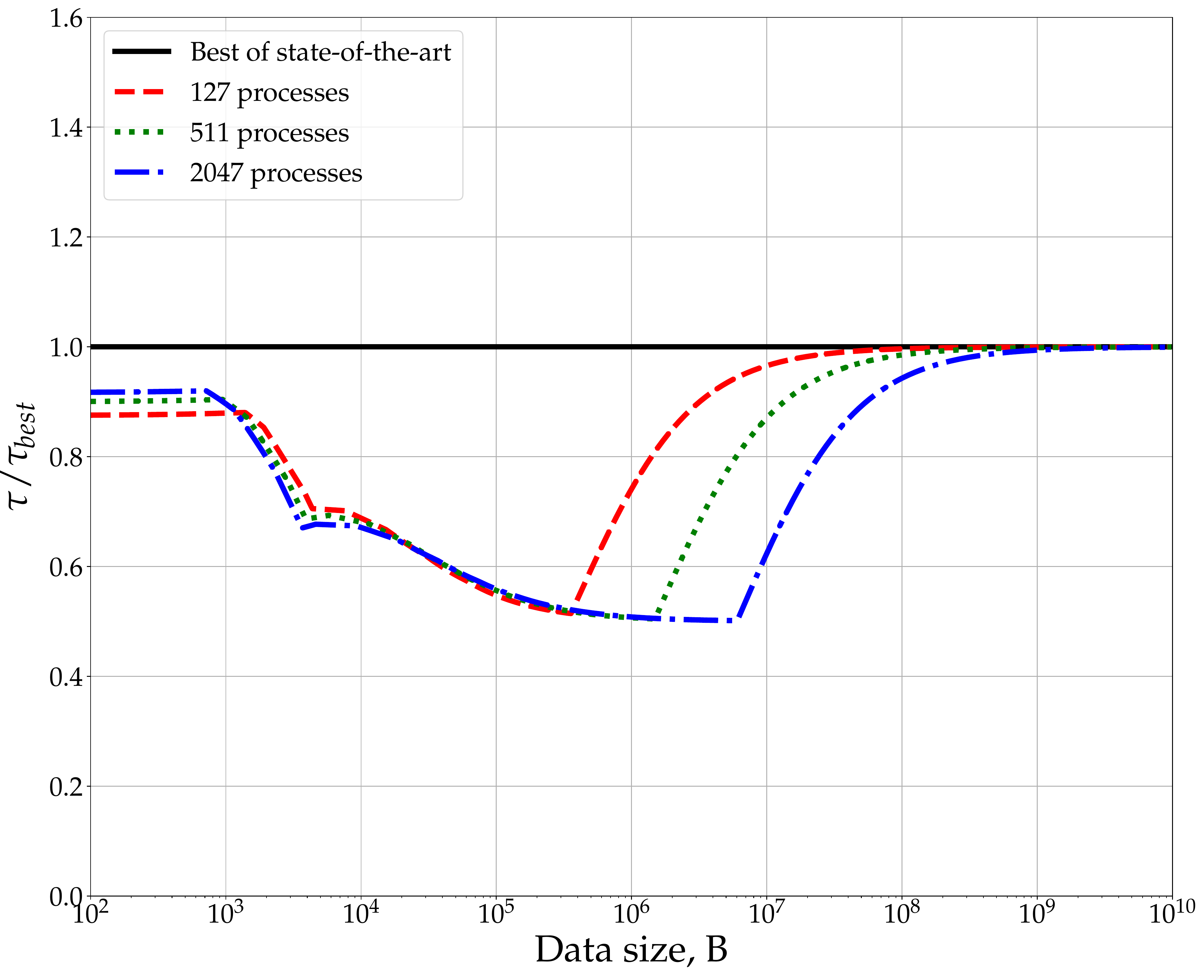}
	\caption{The relationship between the time estimation for the proposed algorithm calculated by \eqref{eq:bo_cost}, \eqref{eq:intemediate_cost}, \eqref{eq:lo_cost} and the best estimation for state-of-the-art algorithms: Ring, Recursive Halving and Recursive Doubling. The proposed algorithm demonstrates the biggest speed-up for the intermediate data sizes where trading-off bandwidth for latency matters. For very large data sets the advantage over the bandwidth-optimal Ring algorithm becomes negligible. Each curve has two discontinuity points which correspond to switching between state-of-the-art algorithms - the first from Recursive Doubling to Recursive Halving and second from Recursive Halving to Ring. }
	\label{fig:estimation-new-to-existing}
\end{figure}

The main goal of this paper is to introduce the novel approach for communication description and the \textit{Allreduce} algorithm with better performance than state-of-the-art solutions for non-power-of-two number of nodes. Figure \ref{fig:estimation-new-to-existing} demonstrates an expected ratio between the time estimation for the proposed algorithm $\tau$ using \eqref{eq:bo_cost}, \eqref{eq:intemediate_cost}, \eqref{eq:lo_cost} and the best estimation for state-of-the-art solutions (Ring, Recursive Halving and Recursive Doubling): $\tau_{best}=min(\tau_{RD}, \tau_{RH}, \tau_{Ring})$. All estimations are made for network parameters given in Table \ref{tab:p2p_params}.

\section{Communication model} \label{comm_model}

To model the communication and computation cost of algorithms, we use a cost model as in Chan et al. \cite{Chan07}:
\begin{displaymath}
\tau_{p2p} = \alpha + \beta m + \gamma{m},
\end{displaymath}
where $\alpha$ is the latency term, $\beta$ is the bandwidth term, $m$ is the message size and $\gamma$ is a computation speed. Most of the presented algorithms divide the initial data vector into $P$ parts. For brevity, we assume that all parts are of equal size: $u=m/P$.

We assume a peer-to-peer network with full-duplex channels, so there are no network conflicts for any cyclic communication pattern. We don't consider any effects caused by particular hardware and network protocol since all possible optimizations related to specific hardware can be adopted for the presented algorithm also.

\section{Related work}

General ideas of so-called butterfly algorithms were described in the works of Van de Geijn \cite{Geijn91}, Barnett et al. \cite{Barnett93} and were further developed by Rabenseifner et al. \cite{Rabenseifner04}, Chan et al. \cite{Chan04}, Traff \cite{Traff05} and Thakur et al. \cite{Thakur05} into the bandwidth-optimal Recursive Halving and the latency-optimal Recursive Doubling. It will be shown in the following sections that both algorithms are special cases of the proposed algorithm. Hybrid algorithms described in \cite{Geijn91,Barnett93,Bruck93,Rabenseifner04,Chan04,Chan07} combine bandwidth- and latency-optimized algorithms to smoothly change the total number of steps. The main problem with the presented butterfly algorithms is that they work only for the power-of-two number of processes case (or the power-of-$(k+1)$ case for k-port systems). 

There are several workarounds for the non-power-of-two number of processes problem. The straightforward solution is to reduce the number of processes to the largest power-of-two $P' < P$ by adding preparation and finalization steps \cite{Barnett93,Bruck93}. At the preparation step excess processes send their data to the processes which perform \textit{Allreduce} and receive the result at the finalization step. This solution requires additional $2m$ data to be sent and the number of steps becomes suboptimal in case of Recursive Doubling. Rabenseifner et al. \cite{Rabenseifner04} introduced a 3-2 elimination protocol which can be plugged into the high-level algorithm such as Recursive Halving to deal with a non-power-of-two number of processes it decreases the overhead to $3/2m$. Our solution doesn't require nor preparation nor finalization steps and thus introduce no overhead.

When combination function is commutative data dissemination algorithms were adopted to perform the reduction \cite{Han88,Hensgen88,Walter92}. These algorithms utilize cyclic communication patterns to distribute information across all processes. Further development of this approach can be found in \cite{End16} where authors proposed a way to use pairwise data exchange paying with bandwidth and latency overhead. These approaches doesn't provide an optimal solution in the non-power-of-two case.

A latency-optimal algorithm based on cyclic communication patterns is described in \cite{Barnoy93}. It also works only with commutative combination function but requires $\ceil{\log{P}}$ steps for any number of processes. This algorithm is optimal in terms of bandwidth and latency but introduces unnecessary computational overhead, moreover, it provides only latency-optimal version which is a corner case for our generalized approach. 

Cyclic communication patterns are also used in \textit{Allgather} algorithm proposed by Bruck et al. \cite{TechBruck93,Bruck97} which is a bandwidth-optimal algorithm working for any number of processes. Its application to the \textit{Allreduce} operation was mentioned in \cite{Bruck93} where the structure of \textit{Allgather} steps is used in reversed order for the reduction phase. This gives, as a result, a bandwidth-optimal \textit{Allreduce} algorithm working for any number of processes. This particular algorithm has the same complexity as our bandwidth-optimal version but it requires additional data shift before reduction phase and after the distribution phase. Also no further development made to trade-off bandwidth for latency. 

Ring algorithm is another bandwidth-optimal \textit{Allreduce} algorithm which performs all communications across the virtual cycle (ring) connecting all processes. Ring algorithm was described by Barnett et al. \cite{Barnett93} and further developed by Patarasuk et al. \cite{Patarasuk07,Patarasuk09} to cover tree topology. This algorithm is advantageous when the vector size is huge since it works with data in a more cache-friendly manner than butterfly algorithms described earlier. Nevertheless, it will be shown that Ring algorithm is directly connected to the butterfly algorithms and is also a special case of the proposed approach.

Multiport algorithms for \textit{Allreduce} operation provide a general solution for systems with an arbitrary number of communication channels $k$ between processes. This is a generalization over the single channel case which leads to a lower number of steps equal to  $\ceil{\log_{P}k}$. The main contribution in this direction was made by Bruck et al. \cite{Bruck93,Bruck97} whose algorithms were described earlier. The development of the data dissemination algorithm mentioned earlier is a k-port dissemination algorithm given in \cite{End16} but only for a limited subset of $\{P, k\}$. 

Generalization of the Recursive Doubling algorithm based on usage of multicast messages is described in \cite{Ruefenacht17}. Optimizations for specific topology are discussed in \cite{Barnett93,Chan07,Kohler12}. In \cite{Hasanov17} authors presented a topology-oblivious version of several existing algorithms. 

There is also an adjacent scientific visualization community where Image Composing algorithm is an analog of the \textit{Allreduce} operation. Radix-k algorithm for big vector size and non-commutative reduction function is proposed in \cite{Peterka09,Kendall10}. It utilizes factorization of $P=p_1\cdot{p_2}\cdot...\cdot{p_n}$ which may be represented as a building of virtual hypercube with dimensions equal to factors and performs linear reduction across each dimension.

\section{Basics of group theory}

In this section, we briefly describe the main mathematical operations used in the following sections.

A \textit{group} is a set (collection of objects) $G$, together with a binary operation "$\cdot$" which inputs two objects in $G$ and outputs a third object in $G$ with the following properties:
\begin{itemize}[noitemsep,topsep=0pt]
	\item the operation "$\cdot$" is associative: 
\begin{displaymath}
(a \cdot b) \cdot c = a \cdot (b \cdot c), a,b,c \in G,
\end{displaymath}
	\item there is an identity object $e$ in $G$:
\begin{displaymath}
e \cdot a = a \cdot e = a,\ \forall a \in G,
\end{displaymath}
	\item every object $a \in G$ has an inverse $a^{-1} \in G$ such that:
\begin{displaymath}
a \cdot a^{-1} = a^{-1} \cdot a = e.
\end{displaymath}
\end{itemize}

The operation does not have to be commutative. Groups for which commutativity always holds are called \textit{abelian groups}. The most famous example of groups is a set of integers together with addition operation. Another example is a permutation group - a set of permutations of a given set $M$ together with a composition operation. We are interested in permutation groups of a specific \textit{order}. The \textit{order} of a group is simply the number of elements in this group. If for each pair $x$ and $y$ in $M$ there exists a $g$ in $G$ such that $g \cdot x = y$ then the group is called \textit{transitive}. \textit{Cyclic group} is another particular type of group which elements are combinations of a single element called \textit{generator}. 

When the order is a prime number, only one cyclic permutation group exists (up to isomorphism). However, for composite numbers, more groups can be found. For example, a cyclic group generated by the permutation 
$ c = \left(\begin{smallmatrix}
	0 & 1 & 2 & 3 & 4 & 5 & 6 & 7 \\
	1 & 2 & 3 & 4 & 5 & 6 & 7 & 0
\end{smallmatrix}\right)$ 
or in cyclic notation $c = (0\ 1\ 2\ 3\ 4\ 5\ 6\ 7)$ is shown in the Table \ref{tab:group-8}.a. It is a transitive permutation group of order $8$ as well as the permutation group given in Table \ref{tab:group-8}.b generated by a set of permutations:
\begin{align} \label{eq:self-inverse-8}
H=\{ &h_1 = (0\ 1)(2\ 3)(4\ 5)(6\ 7), \nonumber \\
&h_2 = (0\ 2)(1\ 3)(4\ 6)(5\ 7), \nonumber \\
&h_3 = (0\ 4)(1\ 5)(2\ 6)(3\ 7)\}.
\end{align}
The last one is a specific type of group which elements are self-inverse so composing it with itself produces the identity element.

\begin{table}
	\caption{Examples of permutation groups. a) Cyclic permutation group of order $8$ with the generator $c = (0\,1\,2\,3\,4\,5\,6\,7)$. b) Permutation group of order $8$ generated by a set of permutations $H$ given in \eqref{eq:self-inverse-8}}
	\label{tab:group-8}
	\subfloat {
	\begin{tabular}{ccl}
		\toprule
		$\#$& Generation combination&In cyclic notation\\
		\midrule
		0 & $c^1$ & $(0\ 1\ 2\ 3\ 4\ 5\ 6\ 7)$ \\
 		1 & $c^2$ & $(0\ 2\ 4\ 6)(1\ 3\ 5\ 7)$ \\
		2 & $c^3$ & $(0\ 3\ 6\ 1\ 4\ 7\ 2\ 5)$ \\
		3 & $c^4$ & $(0\ 4)(1\ 5)(2\ 6)(3\ 7)$ \\
		4 & $c^5$ & $(0\ 5\ 2\ 7\ 4\ 1\ 6\ 3)$ \\
		5 & $c^6$ & $(0\ 6\ 4\ 2)(1\ 7\ 5\ 3)$ \\
		6 & $c^7$ & $(0\ 7\ 6\ 5\ 4\ 3\ 2\ 1)$ \\
		7 & $e=c^8$ & $()$ \\
		\bottomrule
	\end{tabular}}
	\subfloat {
	\begin{tabular}{ccl}
		\toprule
		$\#$& Generation combination&In cyclic notation\\
		\midrule
		0 & $h_1$ & $(0\ 1)(2\ 3)(4\ 5)(6\ 7)$ \\
		1 & $h_2$ & $(0\ 2)(1\ 3)(4\ 6)(5\ 7)$ \\
		2 & $h_1 \cdot h_2$ & $(0\ 3)(1\ 2)(4\ 7)(5\ 6)$ \\
		3 & $h_4$ & $(0\ 4)(1\ 5)(2\ 6)(3\ 7)$ \\
		4 & $h_4 \cdot h_1$ & $(0\ 5)(1\ 4)(2\ 7)(3\ 6)$ \\
		5 & $h_4 \cdot h_2$ & $(0\ 6)(1\ 7)(2\ 4)(3\ 5)$ \\
		6 & $h_4 \cdot h_2 \cdot h_1$ & $(0\ 7)(1\ 6)(2\ 5)(3\ 4)$ \\
		7 & $e=h_1^2=h_2^2=h_4^2$ & $()$ \\
		\bottomrule
	\end{tabular}}
\end{table}

\newsavebox{\smlmat}
\savebox{\smlmat}{
	$h = \left(\begin{smallmatrix}
	0 & 1 & 2 & 3 & 4 & 5 & 6 \\
	4 & 5 & 2 & 6 & 1 & 0 & 3
	\end{smallmatrix}\right)$
}

\section{Communication description via permutations}

Moves of the Rubik's cube\texttrademark\ can be represented as permutations which rearrange the set of cube pieces. A composition of several move permutations generates another permutation. The complete set of all possible permutations forms the Cube permutation group. Recently it was shown by \cite{cube20} that every position of the Cube can be solved in 20 moves or less. It means that any two permutations are connected by less than 20 compositions with move permutations.

In our case, we have a set of processes $\{0,1,...,P-1\}$ that may send data to each other. Let any possible bidirectional data exchange between two processes $i$ and $j$ where $i$ sends data to $j$ and vice versa to be a basic move of our "networking cube." Such a move may be described by a simple transposition: $(i\ j),\ i,j \in \{0,1,...,P-1\}$. At this point, we consider only directions of data exchange without any information about which data is sent. A composition of such elementary transpositions generates other permutations which may be more complex and describe communications involving several processes. For example, a composition of $a=(0\ 1)$ and $b=(1\ 2)$ gives $c=a \cdot b=(0\ 1)(1\ 2)=(0\ 1\ 2)$ which describes a cyclic communication pattern where process $0$ sends data to the process $1$ which sends data to $2$ which finally sends data back to $0$. Moreover, the composition $c'=b \cdot a=(1\ 2)(0\ 1)=(0\ 2\ 1)$ describes an opposite cyclic communication where data is sent from $0$ to $2$ then to $1$ and finally back to $0$. 

\begin{figure}
	\centering
	\includegraphics[scale=0.8]{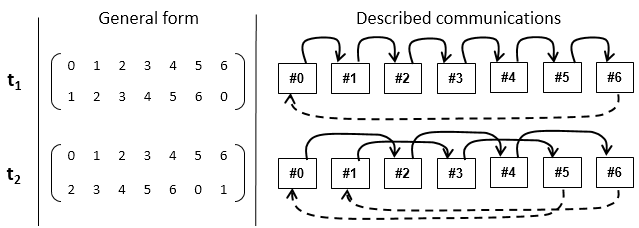}
	\caption{Description of communication patterns via permutation for $P=7$ processes and $T_7$ being a cyclic group with generator $c = (1\ 2\ 3\ 4\ 5\ 6\ 0)$.}
	\label{fig:comm-via-perm}
\end{figure}

The same way as for Rubik's cube\texttrademark\ all possible compositions of permutations for our "networking cube" form a group $W_P$ which contains permutations describing any possible communication pattern between the set of $P$ processes. We are interested in particular subgroups of $W_P$ which have specific properties. The basis of the proposed approach is an abelian permutation group $T_P=\{t_0,t_1,...,t_{P-1}\}$ of order $P$ which acts transitively on the set $\{0,1,...,P-1\}$. Figure \ref{fig:comm-via-perm} shows an example of $T_P$ for the case of $P=7$ processes when it is a cyclic group with generator $c = (1\ 2\ 3\ 4\ 5\ 6\ 0)$ so $t_k = c^k$. In the previous section examples of two suitable groups were presented in Table \ref{tab:group-8}. 

Initial data in the \textit{Allreduce} task is a set of data vectors $\mathbf{V}=\{V_0,V_1,...,V_{P-1}\}$ which are owned by the corresponding processes numbered as $\{0,1,...,P-1\}$. Each vector $V_j$ consists of $P$ parts $V_j=(u_{0,j},u_{1,j},...,u_{P-1,j})$. The whole initial data set can be represented as a matrix (see Figure \ref{fig:initial-distr}.a):
\begin{equation} \label{eq:initial_data}
\mathbf{U}=\{\, u_{i,j} \mid 0 \le i,j < P \,\}
\end{equation}
where columns are vectors $V_j$ and rows are data elements with the same index number.

\subsection{Distributed vector}

\begin{figure}
	\centering
	\includegraphics[scale=0.4]{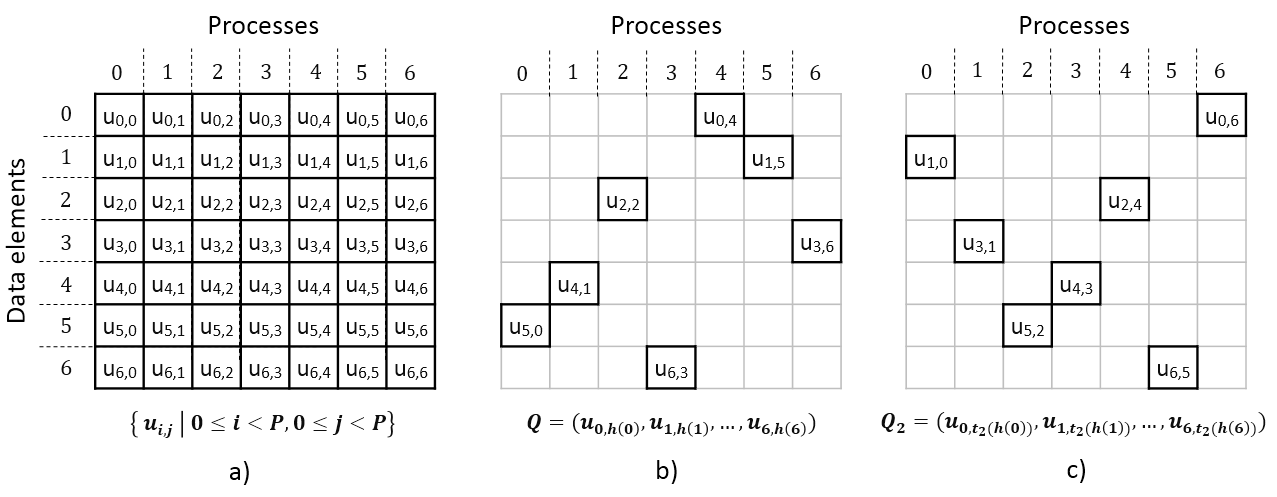}
	\caption{a) Initial data for $P=7$ represented as a matrix of data elements $u_{i,j}$ where columns are owned by the same process and rows are data elements with the same index number. b) Distributed vector $Q$ built using a permutation ~\usebox{\smlmat} defining placements of data elements. c) Distributed vector $Q_2$ built using the same $h$ but its data elements are shifted in accordance with $t_2$. $T_7$ here is a cyclic group with generator $c = (1\ 2\ 3\ 4\ 5\ 6\ 0)$ shown on Figure \ref{fig:comm-via-perm} where $t_k = c^k$. }
	\label{fig:initial-distr}
\end{figure}

Distributed vectors consist of data elements owned by different processes. In our approach we use distributed vectors built in the following way: consider an arbitrary permutation $h$ which acts on a set $\{0,1,...,P-1\}$, then a distributed vector is defined as follows:
\begin{equation} \label{eq:distr_vector_def}
Q=(u_{0,h(0)},u_{1,h(1)},...,u_{P-1,h(P-1)}). 
\end{equation}
An example of a distributed vector \eqref{eq:distr_vector_def} for $P=7$ is shown on Figure \ref{fig:initial-distr}.b. Since the placement of data elements is defined by a permutation then each process owns exactly one data element. 

\subsection{Initial data description}

The initial data \eqref{eq:initial_data} consists of $P^2$ elements $u_{i,j}$, to describe it we need to construct $P$ distributed vectors \eqref{eq:distr_vector_def}:  
\begin{equation} \label{eq:initial_distr_vectors_set}
\bigQ{initial}=\{Q_0,Q_1,...,Q_{P-1}\},
\end{equation}
where $Q_k$ is built using the permutations $t_k$ from the group $T_P$ described earlier:
\begin{equation} \label{eq:initial_distr_vector}
Q_k=(u_{0,t_k(h(0))},u_{1,t_k(h(1))},...,u_{P-1,t_k(h(P-1))}), 
\end{equation}
where $0 \le k < P-1$. An example of $Q_2$ for the case of $P=7$ processes and $T_7$ being a cyclic group with generator $c = (1\ 2\ 3\ 4\ 5\ 6\ 0)$ is shown on Figure \ref{fig:initial-distr}.c.

Since the group $T_P$ is transitive then the group $T'_P = h \cdot T_P$ is also transitive. This means that for any $i$ and $j$ there is a $t'_k$ in $T'_P$ such that $j = i \cdot t'_k$ and since the order of the group is $P$ it has only one such mapping. So the set $\{\, u_{i,t_k(h(i))} \mid 0 \le i,k < P-1 \,\}$ defined by \eqref{eq:initial_distr_vectors_set} and \eqref{eq:initial_distr_vector} is equal to the initial data set \eqref{eq:initial_data}.

Communication operations will change the placement of distributed data vectors, so an additional upper index should be added to data elements to denote its current position:
\begin{displaymath}
Q_k=\Big(u_{0,t_k(h(0))}^{t_k(h(0))},u_{1,t_k(h(1))}^{t_k(h(1))},...,u_{P-1,t_k(h(P-1))}^{t_k(h(P-1))}\Big). 
\end{displaymath}
It is possible to put the permutation $t_k$ out from the upper index of elements giving the notation:
\begin{equation} \label{eq:distr_vector}
Q_k=\mathbf{t_k} q_k, 
\end{equation}
where the current position of data elements in $q_k$ doesn't depend on $k$:
\begin{displaymath}
q_k=\Big(u_{0,t_k(h(0))}^{h(0)},u_{1,t_k(h(1))}^{h(1)},...,u_{P-1,t_k(h(P-1))}^{h(P-1)}\Big)
\end{displaymath}
Examples of $Q_0=\mathbf{t_0} q_0$ and $Q_2=\mathbf{t_2} q_2$ are shown on Figure \ref{fig:initial-distr}.b and Figure \ref{fig:initial-distr}.c respectively.
 
Using the notation \eqref{eq:distr_vector} the initial data set \eqref{eq:initial_distr_vectors_set} can be written as:
\begin{equation} \label{eq:distr_vector_set}
\bigQ{initial}=\mathbf{t_0} q_0 : \mathbf{t_1} q_1 : ... : \mathbf{t_{P-1}} q_{P-1},
\end{equation}
here ":" denotes a concatenation which means that two vectors $t_n \cdot q_n$ and $t_m \cdot q_m$ persist at the same time.

\subsection{Communication operator}

Assume $\mathbf{t_l} \in T_P$ to be a communication operator. Its application to any data vector gives other data vector which data elements change their placement following $\mathbf{t_l}$:
\begin{equation} \label{eq:comm_op_0}
\mathbf{t_l} \cdot  \mathbf{t_k} q_k = \mathbf{t_n} q_k,
\end{equation}
where $\mathbf{t_k}$ stands for the initial placement of vector $q_k$ and its placement after communication is defined by $\mathbf{t_n} = \mathbf{t_l} \cdot \mathbf{t_k}$. 

Since $\mathbf{t_k} q_k$ is a distributed vector the communication is performed by all $P$ processes in parallel, so each process sends one local data element and receives the remote one. Thus all network resources are occupied and it is not possible to increase network utilization performing two communications at once. 

\subsection{Combination of distributed data vectors}

In the \textit{Allreduce} task only data elements with the same indexes can be combined. The result of the reduction phase for a single $i^{th}$ data element:
\begin{displaymath}
u_{i,\Sigma} = u_{i,0} \oplus u_{i,1} \oplus ... \oplus u_{i,P-1}.
\end{displaymath}

Two distributed vectors $q_n$ and $q_m$ can be combined together if they have identical placement, since their data elements with the equal indices are placed on the same processes:
\begin{equation} \label{eq:distr_vectors_combination}
\mathbf{t} q_n \oplus \mathbf{t} q_m = \mathbf{t} \cdot ( q_n \oplus q_m ) = \mathbf{t} \cdot ( q_{n+m} ),
\end{equation}
where $i^{th}$ data element of resulting vector $q_{n+m}$:
\begin{displaymath}
u_{i,n+m}^{h(i)} = u_{i,t_n(h(i))}^{h(i)} \oplus u_{i,t_m(h(i))}^{h(i)},\ 0 \le i < P-1.
\end{displaymath}

As for the communication the combination of two distributed vector is performed by $P$ processes in parallel and each process combines only two local data elements.

\section{Allreduce as permutation composition}

Now when we defined distributed data vectors and basic operations on them, we can start with \textit{Allreduce} description based on the permutation operations. A straightforward solution for the reduction phase is to place all vectors \eqref{eq:distr_vector_set} in the same way and combine them one-by-one. On each step $i$ the following communication operator is applied to the $i^{th}$ data vector:
\begin{equation} \label{eq:reduce_dir_perms}
\mathbf{t_{i \rightarrow 0}} = \mathbf{t_0} \cdot \mathbf{t_i^{-1}},\ 0 \le i < P-1.
\end{equation}

After the communication the reduction operation is performed:
\begin{equation} \label{eq:reduce_step_general}
\mathbf{t_0} q''=\mathbf{t_0} q' \oplus  (\mathbf{t_{i \rightarrow 0}} \cdot \mathbf{t_i} q_i) =  \mathbf{t_0} q' \oplus  (\mathbf{t_0} \cdot \mathbf{t_i^{-1}} \cdot \mathbf{t_i} q_i) = \mathbf{t_0} (q' \oplus q_i),\ 0 \le i < P-1,
\end{equation}
where $q'$ and $q''$ are intermediate combination result. The complete reduction result can be represented as:
\begin{equation} \label{eq:reduce_general}
\mathbf{t_0} q_\Sigma = \mathbf{t_0} q_0 \oplus (\mathbf{t_{1 \rightarrow 0}} \cdot \mathbf{t_1} q_1) \oplus ... \oplus (\mathbf{t_{(P-1) \rightarrow 0}} \cdot \mathbf{t_{P-1}} q_{P-1}) = \mathbf{t_0} \big( q_0 \oplus q_1 \oplus ... \oplus q_{P-1} \big).
\end{equation}

Since only one communication operation can be performed at a time, the straightforward solution requires $P-1$ steps. On each step, the data defined by $\mathbf{t_i} q_i$ is sent to directions defined by the communication operator \eqref{eq:reduce_dir_perms} and combined in accordance with \eqref{eq:reduce_step_general}.

During the distribution phase, reversed communications are performed defined as:
\begin{equation} \label{eq:distr_dir_perms}
\mathbf{t_{0 \rightarrow i}} = \mathbf{t_{i \rightarrow 0}^{-1}}.
\end{equation}

No combination is performed, so each step produces a duplicate of the vector $q_\Sigma$ with different placement. The final result is:
\begin{equation} \label{eq:distr_general}
\bigQ{final}=\mathbf{t_0} q_\Sigma\ :\ \mathbf{t_{0 \rightarrow 1}} \cdot \mathbf{t_0} q_\Sigma\ :\ ...\ :\ \mathbf{t_{0 \rightarrow (P-1)}} \cdot \mathbf{t_0} q_\Sigma = \mathbf{t_0} q_\Sigma\ :\  \mathbf{t_1} q_\Sigma\ :\ ...\ :\ \mathbf{t_{P-1}} q_\Sigma.
\end{equation}

The distribution phase also requires $P-1$ communication steps, so the total number of steps in the straightforward implementation of \textit{Allreduce} is $2(P-1)$, the number of transferred data is $2(P-1) \cdot u$, and the number of computations is $(P-1) \cdot u$, giving the time complexity:
\begin{equation} \label{eq:naive_cost}
\tau_{naive}=2(P-1)\alpha + 2(P-1) u \cdot \beta + (P-1) u \cdot \gamma.
\end{equation}

Now let's consider a special case when $T_P$ is a cyclic group with a generator $\mathbf{t}$. In this case $\mathbf{t_i} = \mathbf{t^i}$ and communication operator \eqref{eq:reduce_dir_perms} can be changed to $\mathbf{t}$. The reduction step \eqref{eq:reduce_step_general} becomes:
\begin{displaymath}
\mathbf{t^{i+1}} q''=\mathbf{t^{i+1}} q_{i+1} \oplus  (\mathbf{t} \cdot \mathbf{t^i} q') = \mathbf{t^{i+1}} (q_{i+1} \oplus q'),\ 0 \le i < P-1,
\end{displaymath}
and the reduction result \eqref{eq:reduce_general} takes the form:
\begin{equation} \label{eq:ring_reduce}
\mathbf{t^{P-1}} q_\Sigma=\mathbf{t^{P-1}} q_{P-1} \oplus \mathbf{t} \cdot ( \mathbf{t^{P-2}} q_{P-2} \oplus \mathbf{t} \cdot ( ... \oplus \mathbf{t} \cdot (\mathbf{t^0} q_{0}))) = \mathbf{t^{P-1}} \cdot (q_0 \oplus q_1 \oplus ... \oplus q_{P-1}).
\end{equation}

During the distribution phase the reduction result \eqref{eq:ring_reduce} is distributed using the same communication operator $\mathbf{t}$ in $P-1$ steps giving $\bigQ{final}$. 

\begin{figure}
	\centering
	\begin{minipage}{0.45\textwidth}
		\centering
		\includegraphics[width=\linewidth]{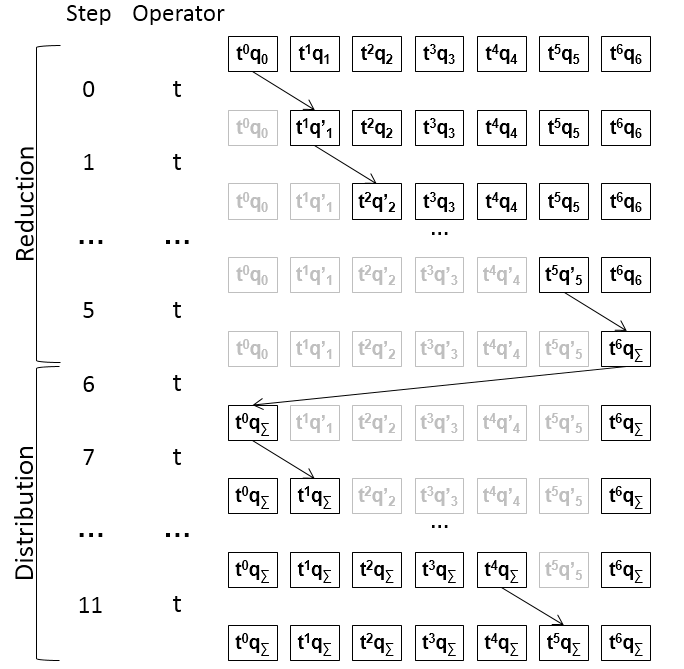}
		\caption{Ring algorithm schedule for $P=7$. Ring algorithm requires $2(P-1)$ steps to complete \textit{Allreduce} with the same communication cycle $t$ used on each step. $q'$ stands got an intermediate reduction result.}
		\label{fig:example-ring-7}
	\end{minipage} \hfill
	\begin{minipage}{0.45\textwidth}
		\centering
		\includegraphics[width=\linewidth]{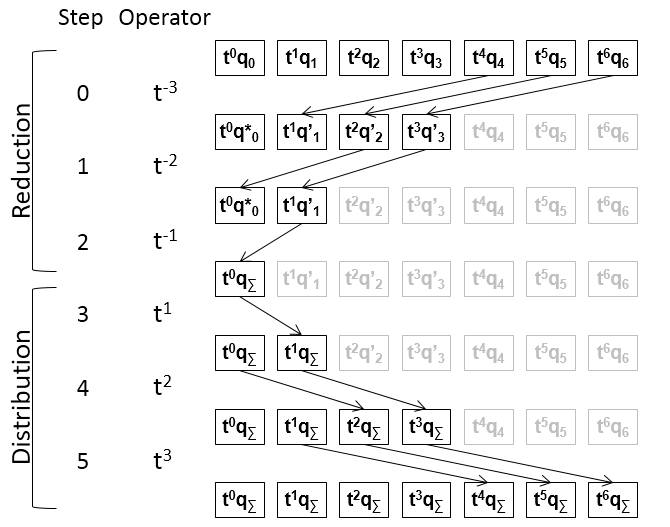}
		\caption{Schedule of the bandwidth-optimal \textit{Allreduce} algorithm for $P=7$. During reduction phase the set of distributed vectors is folded approximately in a factor of 2 on each step giving the reduce result in $\ceil{\log{2}P}$ steps. The distribution phase consists of a reversed set of steps spreading the result vector among all processes.}
		\label{fig:example-bo-7}
	\end{minipage}
\end{figure}

So both reduction and distribution phases can be implemented using the same communication operator $\mathbf{t}$ in $2(P-1)$ steps which is the exact definition of the Ring algorithm. An example schedule of the Ring algorithm for $P=7$ is given on Figure \ref{fig:example-ring-7}.

\setlength{\belowcaptionskip}{-10pt}
\setlength{\belowcaptionskip}{+10pt}

\begin{figure}
	\centering
	\includegraphics[width=\linewidth]{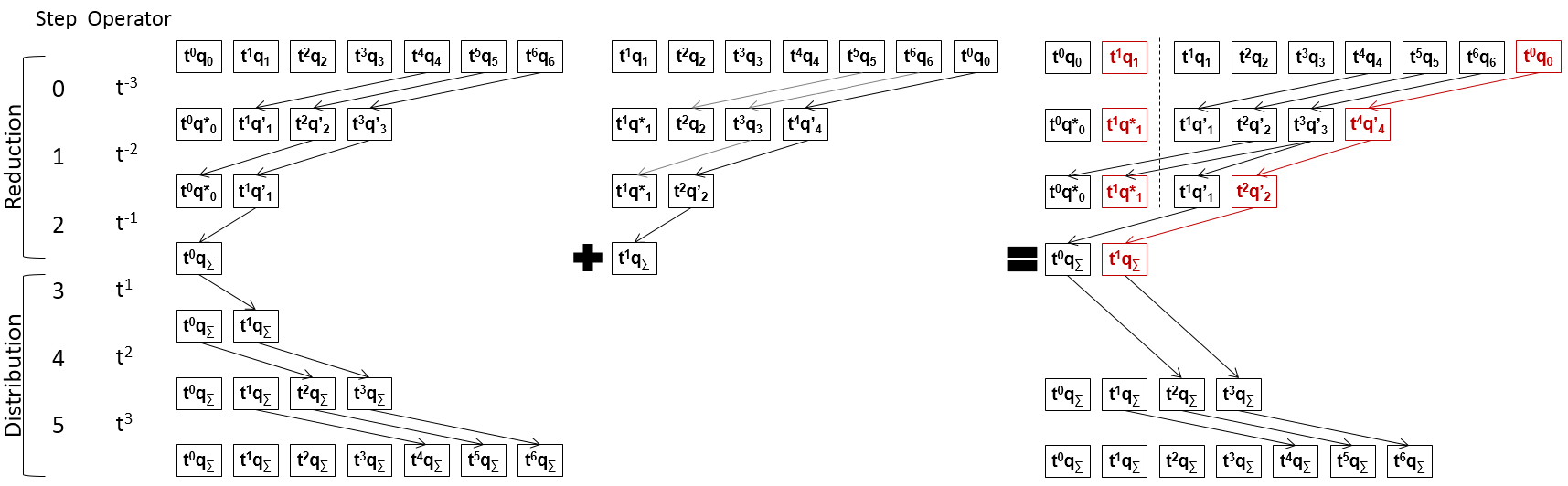}
	\caption{Schedule of the \textit{Allreduce} algorithm for $P=7$ with reduced by $1$ number of steps in the distribution phase. During the reduction phase an additional data vector is transferred on each step to get a bigger result containing of $\comm{0} q_\Sigma$ and $\comm{1} q_\Sigma$. This allows to omit the very first step of the distribution phase. By repeating of such algorithm modification it is possible to decrease the number of steps by the cost of the additionally transferred data.}
	\label{fig:example-bo-7-5steps}
\end{figure}

\section{Bandwidth-optimal algorithm}

The reduction phase can be done much faster if more distributed data vectors are combined on each step. The maximum number of vectors can be combined in one step is $\floor{N_i/2}$ where $N_i$ is the number of separate distributed vectors on step $i$. Such approach requires $\ceil{\log(P)}$ number of steps for the reduction phase. 

The starting set of vectors on $i^{th}$ step is defined by:
\begin{equation} \label{eq:bo_starting_set}
\bigQ{i}=
\begin{cases}
\bigQ{initial}, & \text{if}\ i=0, \\
\comm{0} q_0^* : \comm{1} q'_1 : ... : \comm{N_{i}-1} q'_{N_{i}-1}, & \text{otherwise},
\end{cases}
\end{equation}
where $q'_k$ stands for an intermediate sum of data vectors and $q^*_0$ - is the first data vector which may not participate in the reduction when $N_i$ is odd. The number of distributed vectors $N_{i}$ is:
\begin{equation} \label{eq:bo_starting_set_size}
N_{i}=
\begin{cases}
P, & \text{if}\ i=0, \\
\ceil{N_{i-1}/2}, & \text{otherwise}.
\end{cases}
\end{equation}

Communications on the $i^{th}$ step are defined by the operator:
\begin{equation} \label{eq:bo_dir_perms}
\mathbf{t_{step,i}} = \comm{-\floor{N_{i}/2}},
\end{equation}
which is applied to the set of distributed data vectors:
\begin{equation} \label{eq:bo_tx_dist}
\bigQ{TX,i} = \comm{\ceil{N_{i}/2}} q'_{\ceil{N_{i}/2}} : ... : \comm{N_{i}-1} q'_{N_{i}-1}.
\end{equation}

The part of the set $\bigQ{i}$ that doesn't participate in network communication is defined by:
\begin{equation} \label{eq:bo_rx_dist}
\bigQ{i} \textbackslash \bigQ{TX,i} =  \comm{0} q_0^* : \comm{1} q'_1 : ... : \comm{\ceil{N_{i}/2}-1} q'_{\ceil{N_{i}/2}-1}.
\end{equation}

After the communication is performed $\floor{N_{i}/2}$ vectors will have equal placement and can be combined. The resulting set of vectors after the first step depends on parity of $N_{i}$:
\begin{align}
\bigQ{even,i+1} &= \comm{0} (q_0 \oplus q_{N_{i}/2}) : \comm{1} (q_1 \oplus q_{N_{i}/2 + 1}) : ... : \comm{N_{i}/2-1} (q_{N_{i}/2-1} \oplus q_{N_{i}-1}) = \nonumber  \\ 
&= \comm{0} q'_0 : \comm{1} q'_1 : ... : \comm{\ceil{N_{i}/2}-1} q'_{\ceil{N_{i}/2}-1}, \text{if $N_{i}$ is even}, \label{eq:bo_step_even} \\ 
\bigQ{odd,i+1} &= \comm{0} q_0 : \comm{1} (q_1 \oplus q_{\ceil{N_{i}/2}}) : ... : \comm{\ceil{N_{i}/2}-1} (q_{\ceil{N_{i}/2}-1} \oplus q_{(N_{i}-1)}) = \nonumber  \\ 
&= \comm{0} q_0 : \comm{1} q'_1 : ... : \comm{\ceil{N_{i}/2}-1} q'_{\ceil{N_{i}/2}-1}, \text{otherwise}. \label{eq:bo_step_odd}
\end{align}

At the very last step of the reduction phase $N_{\ceil{log_{2}P}-1}=2$ (this is the only one step for the case $P=2$):
\begin{equation} \label{eq:bo_last_step}
\bigQ{\ceil{log_{2}P}} = \comm{0} q^*_0 : \comm{-1} \cdot \comm{1} q'_1 = \comm{0} (q^*_0 \oplus q'_1) = \comm{0} q_\Sigma .
\end{equation}

The distribution phase is a reversed reduction phase without combination performed on each step. An example schedule for the bandwidth-optimal \textit{Allreduce} for $P=7$ is given on Figure \ref{fig:example-bo-7}. Both reduction and distribution phases take $\ceil{\log(P)}$ while the number of transmitted data and the number of calculations stay the same as for Ring algorithm. So the total cost is:
\begin{equation} \label{eq:bo_cost}
\tau_{bo}=2\ceil{\log(P)} \cdot \alpha + 2(P-1) \cdot u \cdot \beta + (P-1) \cdot u \cdot \gamma .
\end{equation}

The described bandwidth-optimal butterfly \textit{Allreduce} algorithm works for any number of nodes. A similar algorithm can be constructed if use the Bruck's \textit{Allgather} and the same steps in reversed order at the reduction phase as mentioned in \cite{Bruck93}, but this algorithm requires two additional data reorderings: before the reduction phase and after the distribution phase. The proposed bandwidth-optimal algorithm doesn't need data shift after the end of communications since all data vectors appear in the right positions. Moreover, our algorithm is more general since it allows to use any suitable group $T_P$ to vary utilized communication patterns.

When the number of processes is power-of-two and a special group $T_P$ is used (like the group shown in the Table \ref{tab:group-8}.b) then the proposed bandwidth-optimal algorithm becomes equal to the Recursive Halving algorithm. This means that Recursive Halving is a special case of the proposed approach.

\section{Moving towards latency-optimal version}

The idea of changing the number of steps in the butterfly \textit{Allreduce} algorithm appeared in \cite{Geijn91,Barnett93,Bruck93,Rabenseifner04} and was based on the usage of two algorithms at the same time: bandwidth-optimal and latency-optimal. Initially, the reduction phase is started with the bandwidth-optimal algorithm and when the data size to be sent becomes less than some threshold - the latency-optimal algorithm is used. On the distribution phase switching between algorithms is done in the reversed order. The main problem of such hybrid approaches is that they don't work with a non-power-of-two number of processes.

Developing the bandwidth-optimal algorithm discussed in the previous section we get a novel algorithm with ability to change number of steps working for any number of processes. The idea is to get a bigger result on the reduction phase so the distribution phase can take less steps. For example, if we get $t^0 q_\Sigma : t^1 q_\Sigma$ during the reduction phase, we may omit the very first step of the distribution phase (see Figure \ref{fig:example-bo-7-5steps}). We can get the reduction schedule for the $t^1 q_\Sigma$ simply by shifting all vectors in the schedule for $t^0 q_\Sigma$ by $1$ but leave communication operators to be the same. So \eqref{eq:bo_starting_set_size} and \eqref{eq:bo_dir_perms} stay the same while \eqref{eq:bo_starting_set} becomes:
\begin{equation} \label{eq:var_starting_set}
\bigQ{i}'=
\begin{cases}
\bigQ{initial}, & \text{if}\ i=0 \\
\comm{1} q_1^* : \comm{2} q'_2 : ... : \comm{N_{i}} q'_{N_{i}}, & \text{otherwise},
\end{cases}
\end{equation}

Equations \eqref{eq:bo_tx_dist} and \eqref{eq:bo_rx_dist} take the form:  
\begin{align}
&Q'_{TX,i} = \comm{\ceil{N_{i}/2}+1} q'_{\ceil{N_{i}/2}+1} : ... : \comm{N_{i}} q'_{N_{i}}, \\
&Q'_{i-1} \textbackslash Q'_{TX,i} = \comm{1} q^*_1 : \comm{2} q'_2 : ... : \comm{\ceil{N_{i}/2}} q'_{\ceil{N_{i}/2}}.
\end{align}

The step result descriptions \eqref{eq:bo_step_even}, \eqref{eq:bo_step_odd} and \eqref{eq:bo_last_step} change in the similar way:
\begin{align}
Q'_{even,i} &= \comm{1} (q^*_1 \oplus q'_{N_{i}/2+1}) : \comm{2} (q'_2 \oplus q'_{N_{i}/2+2}) : ... : \comm{N_{i}/2} (q'_{N_{i}/2} \oplus q'_{N_{i}}), \\
Q'_{odd,i} &= \comm{1} q^*_1 : \comm{2} (q'_2 \oplus q'_{\ceil{N_{i}/2}+1}) : ... : \comm{\ceil{N_{i}/2}} (q'_{\ceil{N_{i}/2}} \oplus q'_{N_{i}}), \\
Q'_{\ceil{log_{2}P}-1} &= \comm{1} (q^*_1 \oplus q'_2) = \comm{1} q_\Sigma.  
\end{align}

Now it is possible to get a difference between the original equations and new ones. We are interesting in additional data to be sent:
\begin{equation}
Q'_{TX,i} \textbackslash Q_{TX,i} = \comm{N_{i}} q'_{N_{i}},
\end{equation}
and additional data to be reduced:
\begin{align}
Q'_{even,i} \textbackslash Q_{even,i} &= \comm{1} (q^*_1 \oplus q'_{N_{i}/2+1}) : \comm{N_{i}/2} (q'_{N_{i}/2} \oplus q'_{N_{i}}), \\
Q'_{odd,i} \textbackslash Q_{odd,i} &= \comm{1} q*_1 : \comm{\ceil{N_{i}/2}} (q'_{\ceil{N_{i}/2}} \oplus q'_{N_{i}}), \\
Q'_{\ceil{log_{2}P}-1} \textbackslash Q_{\ceil{log_{2}P}-1} &= \comm{1} (q^*_1 \oplus q'_2) = \comm{1} q_\Sigma. 
\end{align}

So adding one data vector $\comm{1} q_\Sigma$ to the result of the reduction phase increases the number of exchanged data by $u$ on each step and also the number of computations by $2u$ for steps when $N_{i}$ is even and by $u$ otherwise. An example of \textit{Allreduce} algorithm schedule for $P=7$ with a reduced number of steps is given in Figure \ref{fig:example-bo-7-5steps}.

To reduce the number of steps further it is required to double the number of resulting data vectors, so the amount of additional data to be sent is $(2^{r}-1) \ceil{\log(P)} \cdot u \cdot \beta$, where $r$ is the number of removed steps from the distribution phase. The number of additional calculations is also doubled, in the worst case when $N_{i}$ is even in $\ceil{\log(P)}-2$ steps the number of additional computations is $(2^{r}-1) (2\ceil{\log(P)}-2) \cdot u \cdot \gamma$. The resulting complexity for the algorithm with an intermediate number of steps:
\begin{align} 
\tau_{i} &= (2\ceil{\log(P)} - r) \cdot \alpha + \nonumber \\
&+ \big( 2(P-1) + (2^{r}-1) (\ceil{\log(P)}-1) \big)  \cdot u \cdot \beta + \nonumber \\
&+ \big( (P-1) + (2^{r}-1) (2\ceil{\log(P)}-2) \big) \cdot u \cdot \gamma, \label{eq:intemediate_cost}
\end{align}
where $0 \le r < \ceil{\log(P)}$. The latency-optimal version ($r = \ceil{\log(P)}$) considered separately in the next subsection.

It is possible to analytically determine the optimal number of steps for the proposed algorithm by finding a minimum of \eqref{eq:intemediate_cost}:
\begin{equation} \label{eq:opt_number_of_steps}
r = \log\Big( \dfrac{\alpha}{m(\beta + 2\gamma)} \Big) + \log\Big( \dfrac{P}{(\log(P) - 1)ln2} \Big)
\end{equation}

The resulting expression consists of two parts: the first one depends on network parameters and the initial data size $m$, the second one depends only on the number of processes. Usage of \eqref{eq:opt_number_of_steps} allows calculating the optimal number of steps based on the estimated network parameters $\alpha, \beta$ and calculation speed $\gamma$.

When the number of processes is power-of-two and a special group $T_P$ is used (like the group shown in the Table \ref{tab:group-8}.b) then the proposed latency-optimal algorithm becomes equal to the Recursive Doubling algorithm which means that Recursive Doubling is a special case of the proposed approach.

\section{Latency-optimal version}

Latency optimal version of the algorithm is a corner case of the version with intermediate number of steps where the reduction phase ends with complete \textit{Allreduce} result. On each step all data vectors are sent so \eqref{eq:bo_tx_dist} becomes: 
\begin{equation}
Q''_{TX} = \comm{0} q'_{0} : \comm{1} q'_{1} : ... : \comm{P-1} q'_{P-1},
\end{equation}
and \eqref{eq:bo_rx_dist}:
\begin{align}
Q''_{i-1} \textbackslash Q''_{TX} =\ \comm{0} q^*_0 : \comm{1} q^*_1 : ... : \comm{P-1} q^*_{P-1} : \ \comm{0} q'_0 : \comm{1} q'_1 : ... : \comm{P-1} q'_{P-1}
\end{align}

Remember that $q^*$ appears at the step when the number of distributed data vectors is odd and reduction does not involve all vectors which leads to the heterogeneous intermediate result. So $q^*$ and $q'$ cannot be combined even if they have the same placement. The step result descriptions \eqref{eq:bo_step_even}, \eqref{eq:bo_step_odd} and \eqref{eq:bo_last_step} take the form:
\begin{align}
&Q''_0 =\comm{0} q^*_0 : \comm{1} q^*_1 : ... : \comm{P-1} q^*_{P-1} : \comm{0} (q_0 \oplus q_{\ceil{P/2}-1}) : \comm{1} (q_1 \oplus q_{\ceil{P/2}}) : ... : \comm{P-1} (q_{P-1} \oplus q_{\ceil{P/2}-2}), \\
&Q''_{even,i} = \comm{0} (q^*_0 \oplus q'_{N_{i}/2}) : \comm{1} (q^*_1 \oplus q'_{N_{i}/2 + 1}) : ... : \comm{P-1} (q^*_{P-1} \oplus q'_{N_{i}/2-1}) : \nonumber \\
& \qquad \qquad : \comm{0} (q'_{0} \oplus q'_{N_{i}/2}): \comm{1} (q'_{1} \oplus q'_{N_{i}/2 + 1}): ... : \comm{P-1} (q'_{P-1} \oplus q'_{N_{i}/2-1}), \\
&Q''_{odd,i} = \comm{0} q^*_0 : \comm{1} q^*_1 : ... : \comm{P-1} q^*_{P-1} : \nonumber \\
& \qquad \qquad : \comm{0} (q'_{0} \oplus q'_{\ceil{N_{i}/2}-1}): \comm{1} (q'_{1} \oplus q'_{\ceil{N_{i}/2}}): ... : \comm{P-1} (q'_{P-1} \oplus q'_{\ceil{N_{i}/2}-2}), \\
&Q'_{\ceil{log_{2}P}-1} = \comm{0} (q^*_0 \oplus q'_1) : \comm{1} (q^*_1 \oplus q'_2) : ... : \comm{P-1} (q^*_{P-1} \oplus q'_0) = \comm{0} q_\Sigma : \comm{1} q_\Sigma : ... : \comm{P-1} q_\Sigma. 
\end{align}

The distribution phase, in this case, is completely omitted. The resulting complexity for the latency-optimal version in the worst case (when $N_{i}$ is even in $\ceil{\log(P)}-2$ steps) is:
\begin{equation} \label{eq:lo_cost}
\tau_{lo}= \ceil{\log(P)} \cdot \alpha + P \ceil{\log(P)}  \cdot u \cdot \beta + P (2\ceil{\log(P)}-2) \cdot u \cdot \gamma.
\end{equation}

\section{Experimental results}

We performed experiments on our cluster of 8 machines with 20 physical cores each connected by 10 GE through the network switch. MPI processes were mapped uniformly to cluster nodes but to achieve properties close to a peer-to-peer network ranks of the processes were randomly shuffled. 

A comparison was made between the proposed algorithm from one side and OpenMPI on the other side. OpenMPI implementation of \textit{Allreduce} utilizes two separate algorithms - Recursive Doubling for data sizes less than 10 KB and Ring for bigger data sets. To have a complete picture we implemented Recursive Halving algorithm which works better on data of medium size. 

To calculate an optimal number of steps for the proposed algorithm simple measurements were performed to estimate latency, bandwidth and computation speed. The estimated values can be found in Table \ref{tab:p2p_params}.

\begin{figure}
	\centering
	\begin{minipage}{0.45\textwidth}
		\centering
		\includegraphics[scale=0.15]{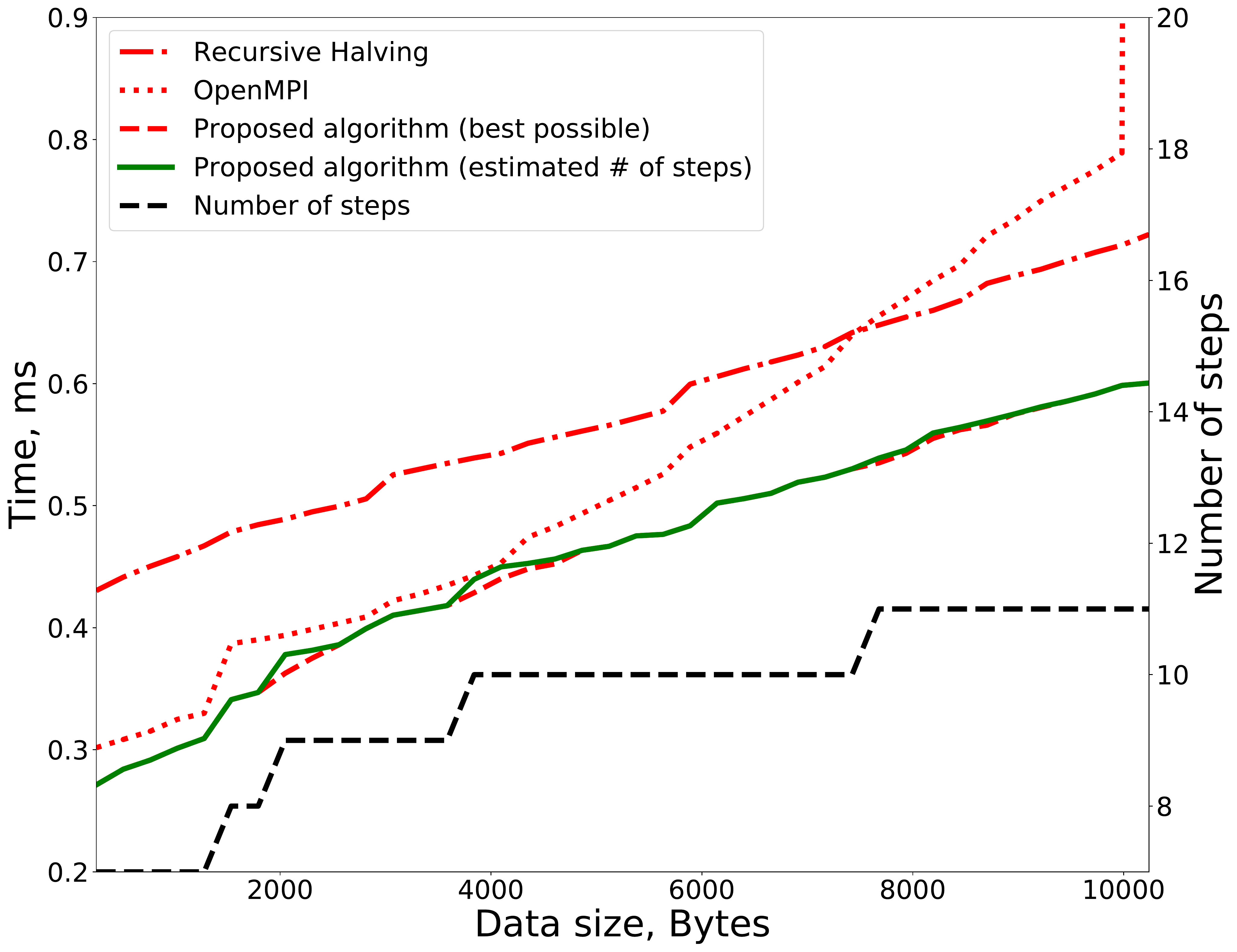}
		\caption{A comparison of the proposed algorithm with OpenMPI and Recursive Halving on small data sets for $P=127$. }
		\label{fig:existing-to-new-comp-smallest}
	\end{minipage} \hfill
	\begin{minipage}{0.45\textwidth}
		\centering
		\includegraphics[scale=0.15]{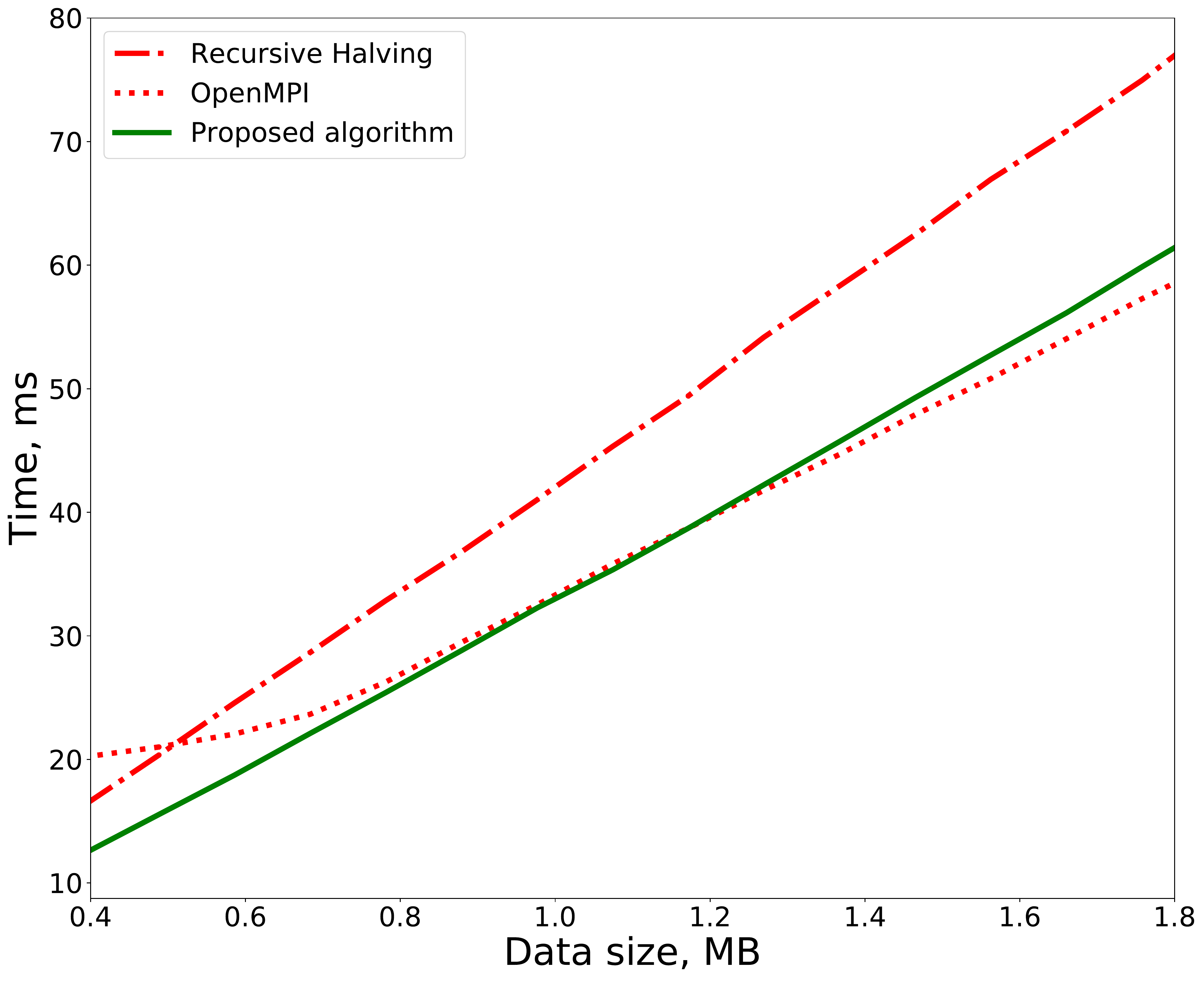}
		\caption{A comparison of the proposed algorithm with OpenMPI and Recursive Halving on big data sets for $P=127$. }
		\label{fig:existing-to-new-comp-big}
	\end{minipage}
	
	\begin{minipage}{0.45\textwidth}
		\centering
		\includegraphics[scale=0.15]{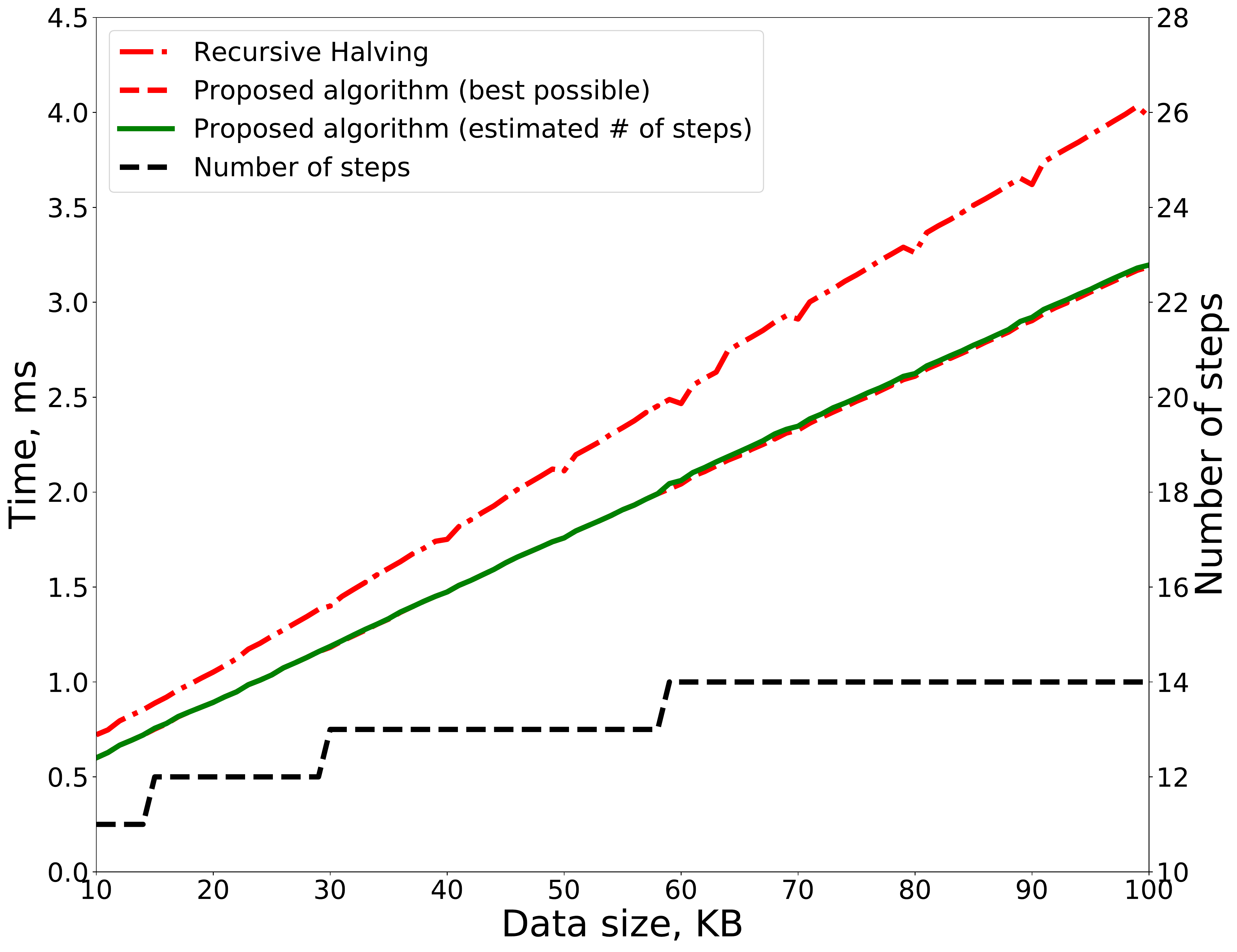}
		\caption{A comparison of the proposed algorithm with Recursive Halving on medium data sets for $P=127$. }
		\label{fig:existing-to-new-comp-small}		
	\end{minipage} \hfill
	\begin{minipage}{0.45\textwidth}
		\centering
		\includegraphics[scale=0.15]{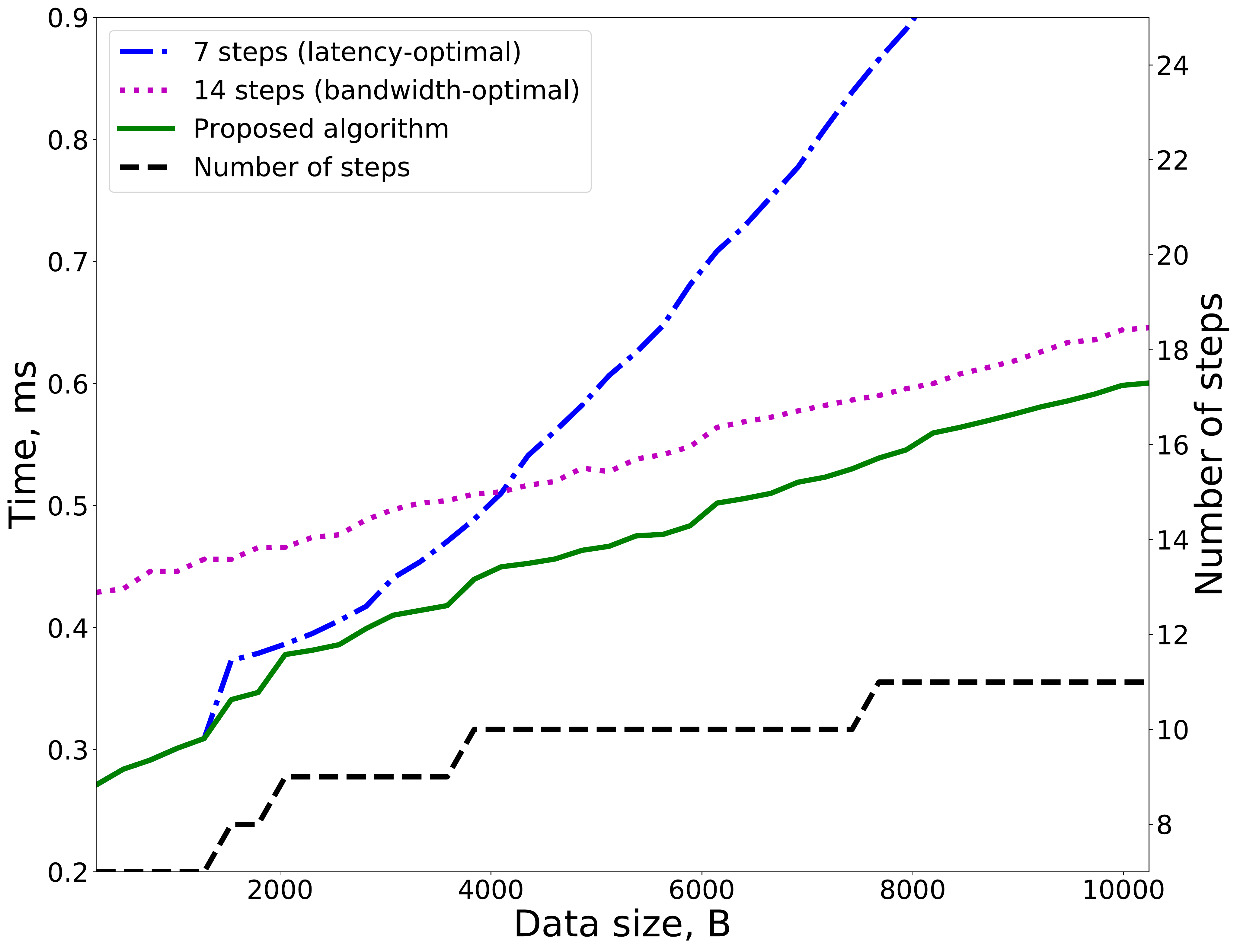}
		\caption{A comparison of different versions of the proposed algorithm for $P=127$. }
		\label{fig:diff-versions-comp}
	\end{minipage}
	
	\begin{minipage}{0.45\textwidth}
		\centering
		\includegraphics[scale=0.15]{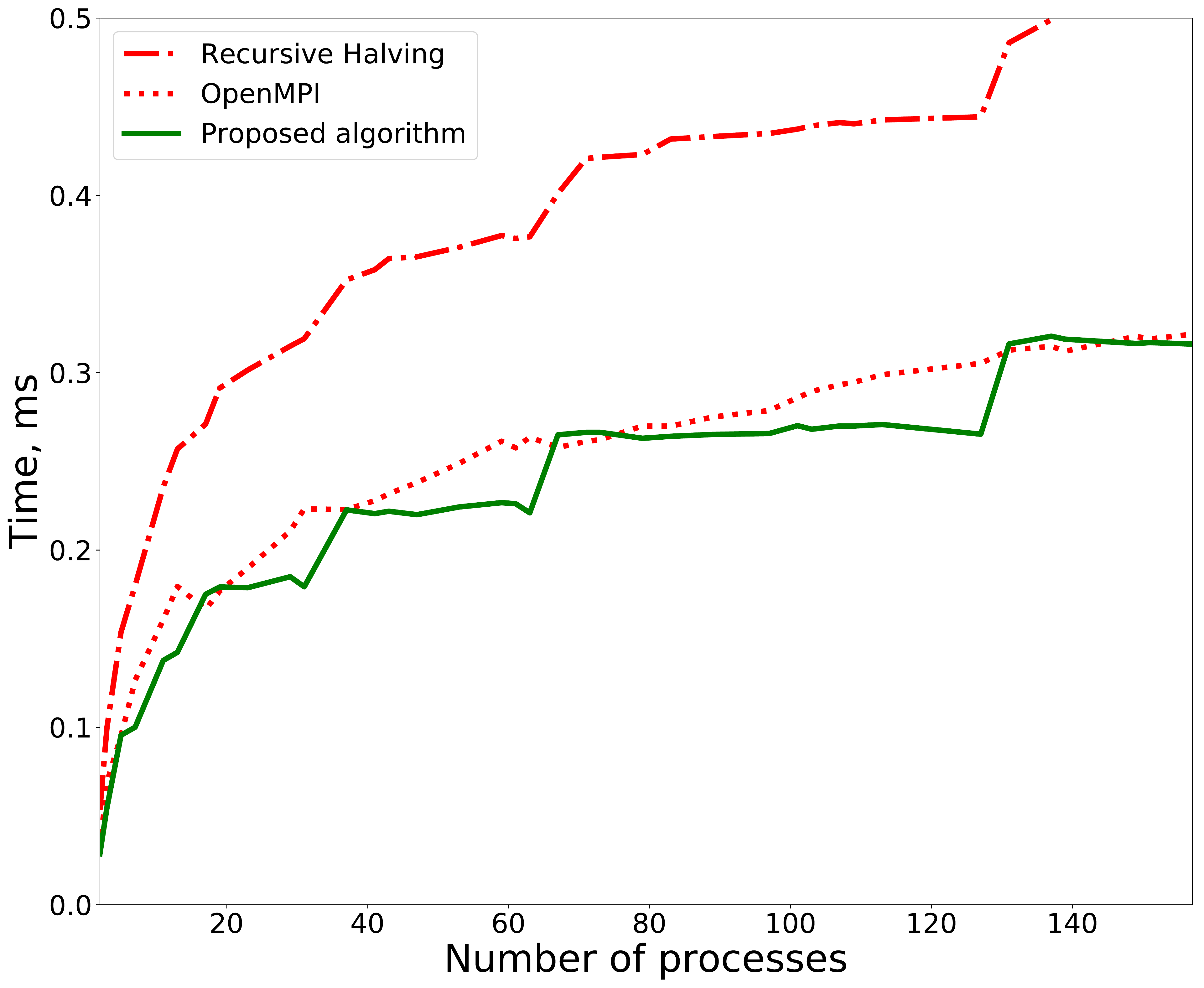}
		\caption{A comparison of \textit{Allreduce} algorithms depending on the number of processes for data size $m=425$ Bytes. }
		\label{fig:existing-to-new-comp-nodes-smallest}
	\end{minipage} \hfill
	\begin{minipage}{0.45\textwidth}
		\centering
		\includegraphics[scale=0.15]{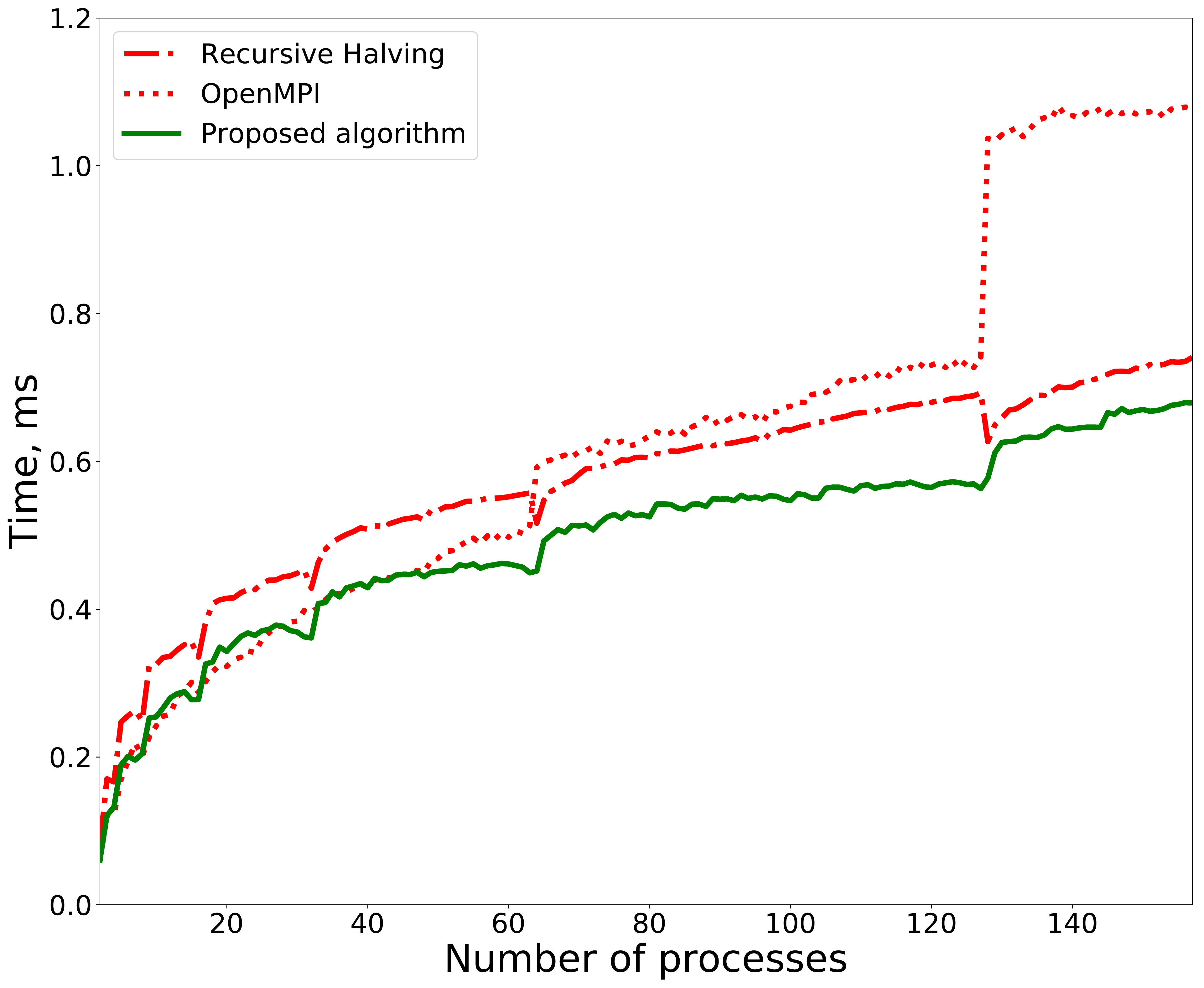}
		\caption{A comparison of \textit{Allreduce} algorithms performance depending on the number of processes for data size $m=9$ KBytes. }
		\label{fig:existing-to-new-comp-nodes-medium}
	\end{minipage}
	
\end{figure}

Figure \ref{fig:existing-to-new-comp-smallest} shows a comparison of the proposed algorithm with OpenMPI and Recursive Halving algorithms performed for small data sizes and the number of processes $P=127$. Performance degradation of OpenMPI at 10 KB caused by switching from Recursive Doubling to Ring algorithm. The proposed algorithm shows better performance even for the estimated number of steps calculated using the network parameters from Table \ref{tab:p2p_params}. There are two reasons: firstly the proposed algorithm has no bandwidth overhead caused by the reduction of the number of processes to the nearest power-of-two, and secondly, it may change the number of steps to better fit current data size. The best possible result for the proposed algorithm (red dashed line) corresponds to the case when we know the exact optimal number of steps. The provided comparison demonstrates that the estimated number of steps fits well with experimental data.

Figure \ref{fig:existing-to-new-comp-small} shows a comparison of the proposed algorithm with Recursive Halving algorithm made for medium data sizes and the number of processes $P=127$. OpenMPI is not presented since it shows too bad performance for given data sizes. The performance gap between the proposed algorithm and Recursive Halving is growing with the size since Recursive Halving has bandwidth overhead caused by the reduction of the number of processes to the nearest power-of-two. 

A comparison for big data sizes is given in Figure \ref{fig:existing-to-new-comp-big}. Ring algorithm (used by OpenMPI) finally beats the proposed one which is due to better cache utilization. However, it is theoretically possible to improve the proposed algorithm in such a way that the number of steps will be bigger than $2\log(P)$ resulting in a smaller size of messages to be sent on each step. This improvement is one of the tasks we are currently working on.

Figure \ref{fig:diff-versions-comp} demonstrates the performance of different versions of the proposed algorithm for $P=127$. The bandwidth-optimal and latency-optimal versions together demonstrate the worst possible performance of the proposed algorithm while the green solid line corresponds to the optimal number of steps for the given data size. The point where bandwidth-optimal and latency-optimal versions intersect shows the biggest benefit from the usage of the flexible number of steps. 

A comparison concerning the number of processes is demonstrated in Figure \ref{fig:existing-to-new-comp-nodes-smallest}. A comparison was made for data size $m=425$ Bytes which is the average size of the data array involved in \textit{Allreduce} \cite{Rabenseifner99}. The number of steps for the proposed algorithm corresponds to the latency-optimal version for all $P$. The performance of Recursive Doubling algorithm isn't changed sharply when the number of processes exceeds power-of-two but instead, it grows very smoothly. It is caused by two reasons: firstly, not all processes participate in the preparation and finalization steps and secondly, the latency for one-way transmission is much smaller than for bidirectional data exchange. In the proposed algorithm all processes participate in bidirectional data exchange on each step which leads to a sharp performance degradation introduced by each additional step. Nevertheless, the proposed algorithm demonstrates much better performance when the number of processes becomes significantly bigger than the nearest power-of-two.

A comparison against the number of nodes for bigger data size $m=9$ KBytes is given in Figure \ref{fig:existing-to-new-comp-nodes-medium}. For big number of processes the proposed algorithm shows better performance even in a power-of-two case since it is able to change the number of steps adopting to the current conditions.

\begin{table}
	\caption{Estimated point-to-point communication parameters for the 10GE cluster used in experiments to calculate the optimal number of steps for the proposed algorithm. }
	\label{tab:p2p_params}
	\begin{tabular}{ccl}
		\toprule
		$\alpha$ & $3 \cdot 10^{-5}$ & $sec$ \\
		$\beta$ & $1 \cdot 10^{-8}$ & $sec / Byte$ \\
		$\gamma$ & $2 \cdot 10^{-10}$ & $sec / Byte$ \\
		\bottomrule
	\end{tabular}
\end{table}

\section{Conclusion}

The described \textit{Allreduce} algorithm is a general solution working for any number of processes and any number of steps between $\log{P}$ and $2\log{P}$. It was shown that widely used algorithms such as Ring, Recursive Halving and Recursive Doubling are special cases of the introduced common approach.  

Our main contributions are:
\begin{itemize}[noitemsep,topsep=0pt]
\item theoretical generalization of existing \textit{Allreduce} algorithms based on group theory;
\item novel algorithm based on the proposed theory which provides latency-optimal and bandwidth-optimal solutions for any number of processes and is able to trade-off latency for bandwidth by changing the number of steps. 
\end{itemize}

While there were some attempts to cover latency-optimal and bandwidth-optimal cases \cite{Barnoy93, TechBruck93, Bruck97} for a non-power-of-two number of processes they are not completely optimal and built with fixed communication patterns. The proposed algorithm provides a complete solution including all versions of \textit{Allreduce} and moreover it is possible to vary utilized communication patterns using different groups $T_P$ which may give a benefit when more complicated network topologies are considered.

A performance comparison was made of the proposed algorithm with existing solutions. The novel algorithm shows significantly better performance on small and medium data sizes. Further optimization of the proposed solution aiming performance improvement for large data sets is possible. For example, it is possible to implement a version of the algorithm which operates with smaller pieces of data but with a bigger number of steps between $2\ceil{\log(P)}$ and $2(P-1)$.

%

%
\bibliographystyle{ACM-Reference-Format}
\bibliography{allreduce-paper}

%


\end{document}